\documentclass[%
 preprint,
 amsmath,amssymb,
 aps,nofootinbib
]{revtex4-1}
\usepackage{dsfont}
\usepackage{graphicx}
\usepackage{dcolumn}
\usepackage{bm}
\usepackage{color}
\usepackage{epsfig}
\usepackage{hyperref}
\usepackage{natbib}
\usepackage{float}
\usepackage{footmisc}
\usepackage{comment}
\usepackage[utf8]{inputenc}
\definecolor{lightred}{rgb}{1,0.5,0.5}
\definecolor{lightgreen}{rgb}{0.5,1,0.5}
\definecolor{lightblue}{rgb}{0.5,0.5,1}
\definecolor{lightcyan}{rgb}{0.5,0.75,0.75}
\definecolor{lightmagenta}{rgb}{0.75,0.5,0.75}
\definecolor{customgreen}{rgb}{0.494,1,0.502}

\newcommand{\mueV}{\mathinner{\mu\mathrm{eV}}}

\newcommand{\eV}{\mathinner{\mathrm{eV}}}

\newcommand{\GeV}{\mathinner{\mathrm{GeV}}}

\begin{document}
\title{Clocking Out Superradiance Limits}
\author{Anubhav Mathur}
\affiliation{Department of Physics and Astronomy, Johns Hopkins University, \\
3400 N. Charles St., Baltimore, MD 21218, USA}
\author{Surjeet Rajendran}
\affiliation{Department of Physics and Astronomy, Johns Hopkins University, \\
3400 N. Charles St., Baltimore, MD 21218, USA}
\author{Erwin H. Tanin}
\affiliation{Department of Physics and Astronomy, Johns Hopkins University, \\
3400 N. Charles St., Baltimore, MD 21218, USA}

\date{\today}

\begin{abstract}
The superradiant instability of black hole space-times has been used to place limits on ultra-light bosonic particles. We show that these limits are model dependent. While the initial growth of the mode is gravitational and thus model independent, the ability to place a limit on new particles requires the mode to grow unhindered to a large number density. Non-linear interactions between the particle and other light degrees of freedom that are mediated through higher dimension operators can damp this growth, eliminating the limit. However, these non-linearities may also destroy a cosmic abundance of these light particles, an attractive avenue for their discovery in several experiments. We study the specific example of the QCD axion and show that it is easy to construct models where these non-linearities eliminate limits from superradiance while preserving their cosmic abundance. 
\end{abstract}

\maketitle

\section{Introduction}
Ultra-light bosonic particles such as axions, axion-like-particles, relaxions and hidden photons have attracted significant attention. These particles are naturally light, emerging as messengers of the deep ultra-violet. They can  solve fundamental problems in particle physics such as the strong CP \cite{Peccei:1977hh, Weinberg:1977ma, Wilczek:1977pj}, hierarchy \cite{Graham:2015cka} and cosmological constant problems \cite{Graham:2017hfr,Graham:2019bfu}. These particles can also be the dark matter of the universe, since they can be copiously produced either in the high temperature universe or during a period of inflation \cite{Preskill:1982cy, Dine:1982ah, Graham:2015rva, Nomura:2015xil, Graham:2018jyp}. This exceptionally strong physics case has motivated a plethora of experiments \cite{Budker:2013hfa, Chaudhuri:2014dla, Graham:2013gfa, Beurthey:2020yuq, Sikivie:2020zpn, Graham:2015ifn, Graham:2017ivz} to directly search for these particles, along with attempts to  impose phenomenological constraints on the viable parameter space of these particles. Many of these constraints arise from astrophysical measurements \cite{Raffelt:2006cw}. This is not a surprise - these particles have  suppressed interactions with the Standard Model and these suppressions are most easily overcome in extreme  astrophysical environments. But, these constraints are not robust. While gross properties of astrophysical environments are reasonably understood, we do not have direct measurements of the micro-physics of these environments \footnote{This is especially true of bounds placed using the superradiance phenomenon in black holes. The black hole information problem poses a serious challenge to confident claims about the physics of black hole horizons. It is plausible that there are classical singularities (or firewalls) at the horizon \cite{Kaplan:2018dqx}, significantly affecting the physics.}.  It is thus plausible that while these constraints may apply to simple models, small modifications can completely remove them \cite{Ahlers:2007rd}.

One interesting class of constraints on light particles  arises from superradiance. In this process, if a particle can be absorbed by a body, when that body rotates, it can become kinematically favorable for that body to emit that particle and spin down. This process can be efficient for bosonic particles around extreme astrophysical bodies such as black holes \cite{Arvanitaki:2009fg, Arvanitaki:2014wva} and neutron stars \cite{Kaplan:2019ako}. In these cases, superradiance can populate the gravitationally bound states of the boson around the star leading to stimulated emission into that state, causing the system to rapidly drain the rotational kinetic energy of the star. Observations of rotating bodies such as black holes and neutron stars can then be used to place limits on the existence of such particles.

In the case of superradiance from black holes, the needed absorptive process is provided by gravitation. Due to the universal nature of gravity, it is often claimed that these bounds are ``model-independent''. This is of course not the case - while it is true that the initial gravitational growth of the superradiant state is model independent, for these bounds to apply, non-linearities in the bound state must be sufficiently small. When non-linear interactions become large, the superradiant amplification can stop, preventing this process from efficiently draining the rotational energy of the system. While some non-linearities may eliminate the superradiance limit on the existence of light particles, they may also significantly change the cosmology of such particles, potentially eliminating them as dark matter candidates. Given the number of experiments searching for ultra-light dark matter over a wide range of parameters, it is thus important to know if non-linearities that would damp superradiance would also preserve the dark matter abundance of such particles. 

In this paper, we show that this is possible. We will specifically study the example of the QCD axion, although our model building approach should be applicable to other scenarios as well. We point out that if the QCD axion has an additional technically natural axion-like interaction with a dark sector that consists of a  light (massless) dark photon and a fermion, these additional interactions can completely eliminate superradiance limits on the axion from rotating black holes, while preserving its cosmic abundance. To achieve this, the axion is assumed to couple more strongly with this dark photon than to QCD. This enhances non-linearities when the superradiant mode begins to grow and thus damps it. The fermions in the dark sector are assumed to have a cosmic abundance - this abundance yields a plasma mass for the hidden photon in the early universe, preventing these non-linearities from affecting the evolution of the dark matter axion. This does not happen around black holes today since these fermions do not have a significant number density around them \footnote{The damping of superradiance due to non-linearities was studied in \cite{Fukuda:2019ewf}. But, they were not concerned with preserving axion dark matter.}. The rest of this paper is organized as follows: in section \ref{sec:model} we present our model and parameters. In section \ref{sec:BH} we show how this parameter space evades super-radiance constraints and in section \ref{s: axiondarkmatter} we show that it preserves axion dark matter. We then conclude in section \ref{sec:conclusions}.

\section{Model}
\label{sec:model}

\begin{figure}
    \centering
    \includegraphics[width=0.5\textwidth]{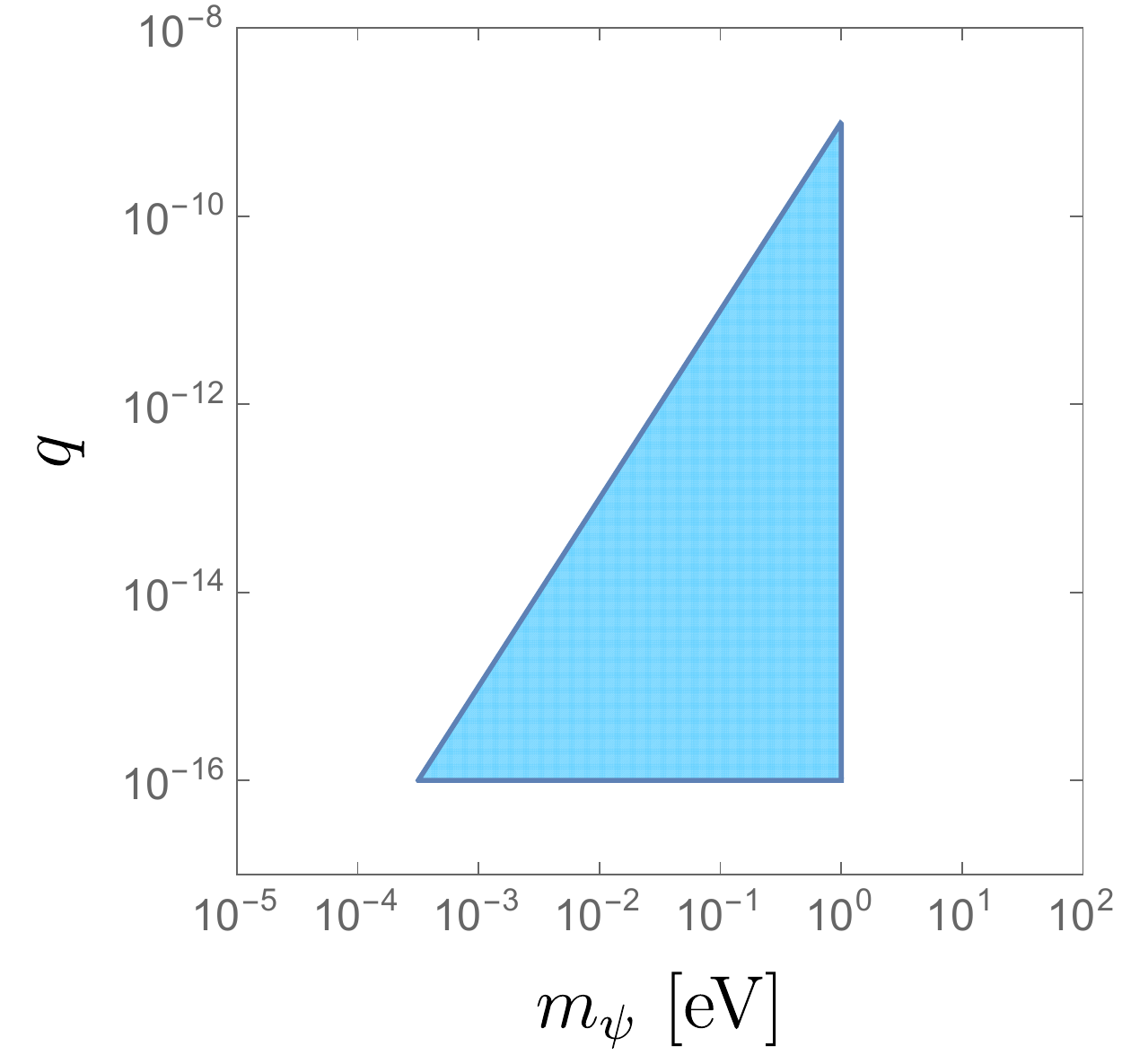}
    \caption{The range of parameter space $(m_\psi,q)$ for $f_a=10^{18}\GeV$ and $f_{\gamma'}=10^{11}\GeV$ satisfying all the constraints and assumptions in Section \ref{s: axiondarkmatter}.}
    \label{fig:par-space}
\end{figure}

We consider the following Lagrangian for the QCD axion $\phi$: 

\begin{equation}
    \mathcal{L} \supset \frac{\phi}{f_{a}} G \tilde{G} + \frac{\phi}{f_{\gamma'}}F'\tilde{F'} + q A'_{\mu} \bar{\psi}\gamma^{\mu}\psi + m_{\psi}\bar{\psi}\psi +\frac{1}{2} m_{a}^2 \phi^2
    \label{model}
\end{equation}
Here $f_a$ is the coupling of the QCD axion to gluons ($G$), $f_{\gamma'}$ is the coupling to a new massless U(1) ($F'$) and $\psi$ are fermions of charge $q$ under $F'$ with mass $m_{\psi}$.  We will be interested in QCD axions in the mass range: 
\begin{equation}
    6\times 10^{-13}\eV\lesssim m_{\rm a}\lesssim 2\times 10^{-11}\eV \label{massspin}
\end{equation}
that are claimed to have been excluded by measurements of the spins of black-holes \cite{Arvanitaki:2014wva}, corresponding to $f_a \approx 10^{18} \text{ GeV}$. Our intent is to simply identify a part of parameter space where this model is self-consistent and achieves our desired goal. A broader analysis may reveal additional parameters where this goal could be achieved - but this is beyond the scope of our work. To that end, we take $f_{\gamma'} \approx 10^{11}$ GeV\footnote{Axions can naturally possess couplings of varying strengths with different gauge groups. This can be accomplished in the so-called aligned axion or clockwork models\cite{Choi:2014rja, Choi:2015fiu, Kaplan:2015fuy}.}, with $q$ and $m_{\psi}$ in the range shown in Figure~\ref{fig:par-space}. In the following sections, we show that we can accomplish our goals with these parameters. 

\section{Axion Mass Bounds from Black hole Superradiance}
\label{sec:BH}

Superradiance is possible whenever the rotation rate ($\omega$) of the black hole is larger than the mass ($m_a$) of the emitted particle. But, this process is efficient only when the Bohr radius of the superradiant mode has significant overlap with the ergoregion ($\sim$ the Schwarzschild radius $r_s$) where superradiant amplification is possible. Together, these conditions imply that superradiance is efficient only when $m_a \approxeq \omega$ and $\omega r_s \approxeq 1$ {\it i.e} the bound requires a nearly extremal black hole and it applies to particles whose masses are close to the rotation rate of the black hole. When these conditions are satisfied, superradiance could extract a significant fraction of the rotational energy $\sim M \left(\omega r_s\right)^2$ of the black hole (whose mass is $M$). This energy is carried by the bosonic field $\phi$. Since efficient superradiance requires the Bohr radius of this field to be $\sim r_s$, we must then have $m_a^2 \phi^2 r_s^3 \sim M \left(\omega r_s\right)^2$. But, since $m_a \approxeq \omega$ and $\omega r_s \approxeq 1$, $\phi \sim M_{pl}$. Given this large field value (and implied number density), limits from superradiance are subject to unknowable non-linear instabilities that could arise due to new physics at very high energies. We now discuss one such instability that arises from the model in \eqref{model}. 

Consider the coupling  
\begin{equation}
    \mathcal{L}_{a\gamma'}= \frac{\phi}{f_{\gamma'}}F_{\mu\nu}^\prime\tilde{F}^{\prime\mu\nu}
\end{equation}
between the axion and the dark photon that we introduced in \eqref{model}. In the presence of such a coupling, efficient dark photon production may occur through processes such as tachyonic instability or parametric resonance (narrow or broad). Regardless of the details of the dark photon production, when enough dark photons are produced, their backreaction to the axion will inevitably thwart the exponential axion growth in order to conserve energy. As we will see, this reduces the maximal Bose enhancement of the axion growth rate, making the whole black hole spindown process effectively slower and weakening the superradiance bounds from black hole spin-mass observations. 

At scales much smaller than the spatial extent $\sim m_a^{-1}$ of the axion bound states\footnote{Here and henceforth we take the ``gravitational fine-structure constant" to be $\alpha\equiv m_a M/M_{pl}^2\sim 1$ because only then is the superradiance growth rate considerably fast.}, the axion field is roughly uniform and can be approximated as
\begin{equation}
    \phi(t)=\phi_0\cos(m_a t)   
\end{equation}
The equation of motion of the circularly polarized dark photon modes $A_k^{\prime\pm}$ then takes the form of the Mathieu equation \cite{Kofman:1997yn,Agrawal:2018vin}
\begin{equation}
    \frac{d^2 A_k^{\prime\pm}}{dz^2}+\left[p_k-2q_k\cos(2z)\right]A_k^{\prime\pm}=0
\end{equation}
with
\begin{equation}
    p_k=\frac{4k^2}{m_a^2},\quad q_k=\pm\frac{2k\phi_0}{f_{\gamma'} m_a},\quad z=\frac{m_a t}{2}
\end{equation}

Exponential growth of the dark photon field may proceed through narrow parametric resonance, broad parametric resonance, and tachyonic instability. In order for the produced dark photons to accumulate, their growth rate must be significantly faster than their escape rate out of the axion cloud, i.e. the inverse of the time it takes for a photon to traverse the size of the axion cloud $\sim m_a^{-1}$ \cite{Hertzberg:2018zte}
\begin{equation}
    \Gamma_{\rm esc}\sim m_a \label{escaperate}
\end{equation}
It is immediately clear that narrow and broad parametric resonance will not be efficient in draining the axion cloud. While the narrow resonance condition $|q_k|\lesssim 1$ for the fastest growing modes (those with $p_k\simeq 1$) translates to the condition $\phi_0\lesssim f_{\gamma'}$, it takes $\phi_0\gg f_{\gamma'}$ for the narrow resonance growth rate $\sim m_a\phi_0/f_{\gamma'}$ to beat $\Gamma_{\rm esc}$. Meanwhile, broad parametric resonance occurs at the rate of $\sim m_a$, which is at best comparable to $\Gamma_{\rm esc}$.

Tachyonic instabilities on the other hand are not similarly inhibited. This instability happens when the squared frequency $\omega_\pm^2=k^2\mp k\frac{\dot{\phi}}{f_\gamma'}$ of one of the circularly polarized dark photon modes becomes negative. Throughout most of an axion oscillation, during which $|\dot{\phi}|\sim m_a\phi_0$, the fastest tachyonic growth occurs for the modes $k\simeq |\dot{\phi}|/(2f_{\gamma'})$ at the rate


\begin{equation}
    \Gamma_{\rm tach}\sim \frac{m_a\phi_0}{f_{\gamma'}}
\end{equation}
As we can see, $\Gamma_{\rm tach}\gg \Gamma_{\rm esc}$ is satisfied\footnote{In order for the dark photon field to grow, the tachyonic growth rate also needs to be much faster than the oscillation rate of the axion, $m_a$, or else the most tachyonic dark photon mode would see a temporally averaged-out axion field. Nevertheless this condition is parametrically similar to $\Gamma_{\rm tach}\gg \Gamma_{\rm esc}$.} when the axion field amplitude $\phi_0$  exceeds $f_{\gamma'}$. This sets an upper bound on the axion amplitude 
\begin{equation}
    \phi_0\lesssim f_{\gamma'}   
\end{equation}
since if $\phi_0$ were to exceed $f_{\gamma'}$, rapid dark photon production would take place, draining energy from the axion field and preventing $\phi_0$ from growing. This suppresses the ability of superradiance to drain rotational energy from the black hole. Suppose the field needed to grow to $\phi_{\text{max}}$ in order to noticeably spin down the black hole \footnote{As per our earlier arguments, qualitatively, $\phi_{\text{max}} \sim M_{pl}$. But there are $\mathcal{O}\left(1\right)$ factors in the estimate - we will thus use the actual field value obtained from the superradiance analysis in our quantitative arguments. }.  In the presence of the coupling with the dark photon, this rate is now limited to $\Gamma_{\text{SR}} \left(\frac{f_{\gamma'}}{\phi_{\text{max}}}\right)^2$, where $\Gamma_{\text{SR}}$ is the unsuppressed superradiance rate in the absence of the axion interaction with the dark photon. Let us now see how this relaxes bounds on axions from superradiance.

\subsection{Relaxation of Bounds from Observed Black Holes}\label{s:possiblebounds}

When superradiance occurs at a rate $\Gamma_{\text{SR}}$, it needs a time $\tau_{\rm BH}$ where
\begin{equation}
    \Gamma_{\rm SR}\tau_{\rm BH}\gtrsim 100 \label{aprioribound}
\end{equation}
in order for the superradiant mode to sufficiently grow and damp the rotation of the black hole. Ref. \cite{Arvanitaki:2014wva} used this condition to derive the axion mass bound \eqref{massspin} from the reported spins, masses, and ages of five stellar mass black holes: M33 X-7, LMC X-1, GRO J1655–40, Cygnus X-1, and GRS 1915+105. They obtained the superradiance rate $\Gamma_{\rm SR}$ using the semianalytical method detailed in \cite{Dubovsky} and adopted
\begin{equation}
    \tau_{\rm BH}=\text{min}\left(\tau_{\rm age},\tau_{\rm Edd}\right)
\end{equation}
where $\tau_{\rm age}$ is the age of the black hole and $\tau_{\rm Edd}=4\times 10^{8}$ years is the Eddington time. This is tantamount to assuming that the black hole accretes at most at the Eddington rate since its birth.  

In order to place limits, one needs measurements of black hole spin. For the purposes of this section, we take the measurements of black hole spin quoted in \cite{Arvanitaki:2014wva} at face value (we comment on the quality of these measurements and the robustness of inferred bounds in section \ref{sec:conclusions}). In the absence of a coupling to dark photons, the axion masses (\ref{massspin}) to which a given black hole is sensitive are simply those values of $m_a$ for which $\Gamma_{\rm SR}(m_a, M)$ satisfies (\ref{aprioribound}). In the absence of an analytic solution for the spectrum of superradiant levels, we follow \cite{Discovering} in using the ``semianalytical" $\Gamma_{\rm SR}$ shown in Fig. 5 of \cite{Dubovsky}, which agrees with analytical formulas under the $r_s m_a \ll \ell $ and WKB approximations, as well as with numerical results.

When the dark photon coupling is added, the maximum rate of rotational energy extracted from the black hole is suppressed by a factor $\left(\frac{f_{\gamma'}}{\phi_{\text{max}}}\right)^2$. Using this suppressed rate, we find the largest value of $f_{\gamma'}$ for which $ \Gamma_{\text{SR}} \tau_{\text{BH}} \lessapprox \text{max}\left[100, \left(\frac{\phi_{\rm max}}{f_{\gamma'}}\right)^2\right]$ for all black holes considered in \cite{Arvanitaki:2014wva}. It suffices to consider the highest superradiant mode $n=1,\,\ell=m=1$. We find that a coupling of $f_{\gamma'} \sim 10^{11} \GeV$ results in the total removal of the superradiance bounds.

The axion to dark photon coupling scale $f_{\gamma'}$ is highly model dependent and can in principle go to much lower values, especially if one invokes the clockwork scenario \cite{Choi:2014rja,Choi:2015fiu,Kaplan:2015fuy}. Further, since $f_{\gamma'}$ couples two dark sectors (namely, axions and dark photons), there do not currently exist any model-independent constraints on its value.

\section{Axion Dark Matter}\label{s: axiondarkmatter}
The arguments of section \ref{sec:BH} show that there are no model independent limits from superradiance on the existence of axions. But,  many experimental methods to discover the axion are aimed at searching for a cosmic abundance of these particles. This has been considered an appealing path since generic initial conditions of the universe end up producing such a cosmic abundance. It is thus interesting to ask if the non-linear interactions introduced in \eqref{model} would also similarly deplete the cosmic abundance of the axion. Indeed, such depletion is possible and it can be used to eliminate the mild fine tuning of the initial axion angle necessary in QCD axion models with $f_a \gtrapprox 10^{12}$ GeV. Since our goal is to simply eliminate superradiance limits while preserving axion dark matter, we present a simple scenario where the cosmic abundance is preserved.

To that end, we add a fermion charged under the dark photon. This fermion is assumed to have a cosmic abundance, giving the dark photon a plasma mass in the early universe. This plasma mass suppresses the cosmological tachyonic instability, preserving axion dark matter. Since it does not accumulate in significant numbers around black holes, the damping of superradiance around black holes is unaltered. We now show a parameter space that accomplishes this goal. 

As shown in \eqref{model}, consider the interactions of the dark photon $A_{\mu}\prime$ to a dark fermion $\psi$: 
\begin{equation}
    \mathcal{L}\supset qA_\mu^\prime\bar{\psi}\gamma^\mu\psi+m_\psi\bar{\psi}\psi
\end{equation}
In the presence of a plasma of these fermions with temperature $T_\psi$ and number density $n_\psi$, the dark photons acquire a plasma mass \cite{Braaten:1993jw}
\begin{equation}
    \omega_p\sim \begin{cases}
    qT_\psi, &T_\psi\gtrsim m_\psi\\
    q\sqrt{n_\psi/m_\psi}, &T_\psi\lesssim m_\psi
    \end{cases}
\end{equation}
Tachyonic instabilities are suppressed as long as
\begin{equation}
    \omega_p \gg \frac{\dot{\phi}}{f_{\gamma'}}
\end{equation}
Before demonstrating a viable parameter space, we make the following simplifying assumption. We will assume that the charge $q$ of the fermion is: 

\begin{equation}
    q\lesssim \frac{m_\psi^2}{m_a f_{\gamma'}}\sim 10^{-9}\left(\frac{m_\psi}{1\eV}\right)^{2}\left(\frac{f_a}{10^{18}\GeV}\right)\left(\frac{10^{11}\GeV}{f_{\gamma'}}\right)
    \label{noSchwinger}
\end{equation}
We do this in order to ignore the complication that the superradiant process around a black hole that produces the dark photon may also simultaneously produce these charged fermions. One may worry that if such charged fermions are produced around the black hole, they could contribute to a residual plasma mass that may suppress the tachyonic instability around the black hole as well. While this is a possibility, further analysis is required to see if this is actually an issue: in the absence of confining forces, it is possible that the produced fermions will simply leave the black hole without creating a plasma that would block the effect. We simply choose to avoid this problem. The condition \eqref{noSchwinger} is equivalent to requiring that there is no Schwinger pair production of the dark fermions from the dark electric field building around the black hole during superradiance {\it i.e.} \eqref{noSchwinger} implies that the dark electric field $E^\prime$, which is at most comparable to the square root of the maximum energy density of the axion field, $m_af_{\gamma'}$, does not exceed the Schwinger limit $E_{\rm crit}^\prime\sim m_\psi^2/q$.

The dark photon and fermion are uncoupled to the Standard Model except through the highly suppressed axion portal. Observational limits on them are thus weak - they are limited by bounds on dark radiation and dark matter. Given this freedom, we make further simplifying assumptions: 

\begin{enumerate}
    \item The dark fermions $\psi$ are in thermal equilibrium among themselves with a temperature
    \begin{equation}
        T_{\psi}=\kappa T
    \end{equation}
    where $T$ is the temperature of the Standard Model sector and $\kappa$ is $O(1)$ but small enough to avoid violating the effective relativistic degrees of freedom $N_{\rm eff}$ bound, $\Delta N_{\rm eff}\lesssim 0.3$ \cite{Cyburt:2015mya}.
    \item $m_\psi\ll 1 \eV$
    \item The dark fermions are thermally decoupled from the dark photons. This forces\footnote{The annihilation rate of the dark fermions $\psi$ to dark photons $\gamma'$ is $\Gamma_{\rm ann}\sim 10^{-3}q^4T\text{ min}\left(1, \frac{T^2}{m_\psi^2}\right)$. When $\psi$ is relativistic, the Hubble rate $H\propto T^2$ scales faster than $\Gamma_{\rm ann}$. Conversely, when it is non-relativistic, $H\propto T^{3/2}$ scales slower than $\Gamma_{\rm ann}$. Correspondingly, the ratio $\Gamma_{\rm ann}/H$ peaks at $T\sim m_\psi$ and requiring $\Gamma_{\rm ann}$ to be greater than the matter-domination expression of $H$ at $T\sim m_\psi$ imposes an upper bound on the dark fermion charge, $q\lesssim  10^{-6}\left(m_\psi/1\eV\right)^{1/8}$.}
    \begin{equation}
        q\lesssim  10^{-6}\left(\frac{ m_\psi}{1\eV}\right)^{1/8} \label{qdecoupled}
    \end{equation}
\end{enumerate}
Assumption 1 ensures that the energy density of $\psi$ is negligible compared to that of the Standard Model radiation when $\psi$ is relativistic, assumption 2 ensures that the energy density of $\psi$ is negligible compared to that of the dark matter when $\psi$ is non-relativistic, and assumption 3 helps ensure that tachyonic instability is blocked throughout cosmic history by preventing Boltzmann suppression of the $\psi$ abundance.

At the beginning of axion oscillation, the requirement that tachyonic instability be blocked amounts to $\dot{\phi}/f_{\gamma '} < \omega_p$, which imposes the constraint
\begin{equation}
    q\gtrsim \frac{m_a f_a\theta_i}{T_{\rm osc}f_{\gamma'}}\sim 10^{-16}\left(\frac{f_a}{10^{18}\GeV}\right)^{1/8}\left(\frac{10^{11}\GeV}{f_{\gamma'}}\right) \label{tachatTosc}
\end{equation}
where $T_{\rm osc}\sim 0.3 \sqrt{m_aM_{pl}}$ is the temperature at which the axion begins to oscillate \cite{Agrawal:2017eqm} and $\theta_i$ is the initial axion angle. As long as $\psi$ is relativistic, the plasma mass $\omega_p\propto T$ decays slower than $\dot{\phi}/f_{\gamma'}\propto T^{-3/2}$, and tachyonic instability remains blocked. Furthermore, thanks to the non-zero chemical potential of $\psi$ that comes with assumption 3, the $\psi$ abundance always follows a Fermi-Dirac distribution with total number density $n_\psi\sim T^3$, even when $T\ll m_\psi$. Consequently, the two quantities of interest, $\dot{\phi}/f_{\gamma'}\propto T^{-3/2}$ and $\omega_p\propto \sqrt{n_\psi}\propto T^{-3/2}$, get Hubble diluted at the same rate when the $\psi$ fermions become non-relativistic and the tachyonic instability continues to be blocked.

Accounting for all the above constraints and assumptions, the remaining parameter space is shown in Figure \ref{fig:par-space}. That said, there is certainly more viable parameter space beyond what we have identified within the boundaries of our simplifying assumptions.

Given the above discussion, the reader may wonder if a dark sector was necessary to accomplish our goals, or if they could have been accomplished with the photon. Much of our discussion applies to the photon as well. The extremely dense plasma in the early universe would inhibit the conversion of axions to photons, preserving axion dark matter. While the photon is better constrained than the dark photon, current limits \cite{Chen:2018} on the coupling would still allow us to eliminate much of the superradiance bound on axions. However, there is a key uncertainty - we do not know the plasma mass (or its spatial distribution) around the black hole. This plasma mass could range between $10^{-11}-10^{-2}\eV$, depending on the details of the plasma environment around the black hole \cite{Sen:2018cjt,Boskovic:2018lkj}. In regions where the plasma mass is small enough, our mechanism will kick in and suppress superradiance. But, absent a better understanding of the spatial distribution of the plasma mass around the black hole, we cannot place robust exclusions.  



\section{Conclusions}
\label{sec:conclusions}
We have shown that superradiance limits on light particles from measurements of black hole spins are model dependent. While the initial growth of the superradiant modes is due to gravity,  the bound requires a large number density (with field values $\sim M_{pl}$) of the new particle around the black hole. This density could trigger instabilities that dampen superradiance. We have also shown that these instabilities can exist only around black holes without affecting the cosmic abundance of axion dark matter. 

The parameter choices we made in this paper were obtained by taking the black hole spin measurements quoted in  \cite{Arvanitaki:2014wva} at face value. However, the spin measurements used in \cite{Arvanitaki:2014wva}  are not conservative. There is inherent astrophysical uncertainty in these measurements. Existing measurements of black holes spins are based on the inference of the innermost stable circular orbit (ISCO) of the accretion disk surrounding the black hole which is expected to be a monotonically decreasing function of the black hole spin. Unfortunately, this method is highly dependent on the astrophysical model used. At this time, there is not even community agreement on the measured spin. This is a critical issue for bounds on superradiance from black hole spin measurements - this process is efficient only for nearly extremal black holes. If the spin was 20-30 percent smaller, the bounds would be significantly different and in some cases entirely absent. At present, it is not clear if the astrophysical model can attain the level of precision needed for these bounds. 

Interestingly, if the actual spin was somewhat lower than the nearly extremal values used in \cite{Arvanitaki:2014wva}, it might be possible to eliminate superradiance limits with just a single dark photon instead of also additionally requiring dark fermions. At lower spin rates, the superradiance rate is significantly slower,  enabling $f_{\gamma'}$ to be higher. This larger value of $f_{\gamma'}$ would deplete the cosmic abundance of the axion - but it is possible that this depletion is at the right level required to avoid over production of axion dark matter, solving the problems of the so-called ``anthropic'' axion window without fine-tuning. In this paper, we focused on demonstrating the model dependence of superradiance limits on axions. Similar limits have also been placed on other scalars and dark photons. It would be interesting to see if similar models can be constructed in these scenarios where non-linearities suppress superradiant growth while preserving a cosmic abundance of these particles. 

\section*{Acknowledgements}
We would like to thank Emanuele Berti, Michael Fedderke, Peter Graham, David E. Kaplan and Julian Krolik for  discussions. S.R.~was supported in part by the NSF under grants PHY-1818899 and PHY-1638509.

\bibliography{references}
\bibliographystyle{unsrt}

\end{document}